          [2001/04/25 1.1 (PWD)]
\documentclass[a4paper,twocolumn]{esapub2005} 
\usepackage{times}

\usepackage{natbib}
\usepackage{graphicx}

\title{Supergiant Fast X-ray Transients:
A common behaviour or a class of objects?}

\author[1,2]{Ignacio Negueruela}
\author[3]{David M. Smith}
\author[1]{Jos\'e Miguel Torrej\'on}
\author[4]{Pablo Reig}
\affil[1]{Departamento de F\'{\i}sica, Ingenier\'{\i}a de Sistemas y
Teor\'{\i}a de la Se\~{n}al, Universidad de Alicante, Apartado 99,
E03080 Alicante, Spain}
\affil[2]{Department of Physics and Astronomy, The Open University,
  Walton Hall, Milton Keynes MK7 6AA, United Kingdom }
\affil[3]{Physics Department \& Santa Cruz Institute for Particle
  Physics, University of California Santa Cruz, 1156 High Street, Santa
  Cruz, CA 95064, U.S.A.}
\affil[4]{IESL (FORTH) \& Physics Department.,University of Crete,  PO Box
  2208, 71003 Heraklion, Crete, Greece}

\def\la{\mathrel{\mathchoice {\vcenter{\offinterlineskip\halign{\hfil
$\displaystyle##$\hfil\cr<\cr\sim\cr}}}
{\vcenter{\offinterlineskip\halign{\hfil$\textstyle##$\hfil\cr
<\cr\sim\cr}}}
{\vcenter{\offinterlineskip\halign{\hfil$\scriptstyle##$\hfil\cr
<\cr\sim\cr}}}
{\vcenter{\offinterlineskip\halign{\hfil$\scriptscriptstyle##$\hfil\cr
<\cr\sim\cr}}}}}

\begin{document}

\keywords{binaries: close --- stars: supergiants -- X-rays: binaries}

\maketitle

\begin{abstract}
{\it INTEGRAL} monitoring of the Galactic Plane is revealing a growing
number of recurrent X-ray transients, characterised by short outbursts
with very fast rise times ($\sim$ tens of minutes) and typical
durations of a few hours. A substantial fraction of these sources
is associated with OB supergiants and hence defines a new class of
massive X-ray binaries, which we call Supergiant Fast X-ray
Transients. Characterisation of the astrophysical parameters of their
counterparts is underway \footnote{Partially based on observations collected
   at the European Southern Observatory, Paranal, Chile (ESO
   077.D-0055,274.D-5010), the Nordic Optical Telescope (NOT) and the
   South African Astronomical Observatory.}. So far, we have found a
   number of late O and 
early B supergiants of different luminosities at a large range of
distances. Nothing in their optical properties sets them apart from
classical Supergiant X-ray Binaries. On the other hand, there is now
rather concluding evidence that persistent supergiant X-ray binaries
also show fast outbursts. This suggests a continuum of behaviours
between typical persistent supergiant systems and purely transient
systems, but offers very little information about the physical causes
of the outbursts.
\end{abstract}

\section{Introduction}

High Mass X-ray Binaries (HMXBs) are X-ray sources composed of an
early-type massive star and an accreting compact object. Most known
HMXBs are Be/X-ray binaries, systems consisting of a neutron star
accreting from the disc around a Be star. Even 
though a few Be/X-ray binaries are persistent weak X-ray sources (with
$L_{{\rm X}}\sim10^{34}\:{\rm erg}\,{\rm s}^{-1}$), the majority are
transients, displaying bright outbursts with typical duration of the
order of several weeks.

The second major class of HMXBs contains early-type supergiants. These
objects are hence known as Supergiant X-ray Binaries (SGXBs). The
compact object is fed by accretion from the strong radiative wind of
the supergiant. These objects are persistent sources, with
luminosities around $L_{{\rm X}}\sim10^{36}\:{\rm erg}\,{\rm s}^{-1}$,
very variable on short timescales, but rather stable in the long
run. About a dozen SGXBs were known before the launch of
{\it INTEGRAL}, most of them having been discovered in the early days of X-ray
astronomy. This low number was generally attributed to a real scarcity
of such systems, as the short duration of the supergiant phase would
result in very short lifetimes. Since the launch of {\it INTEGRAL},
however, the situation has changed
dramatically, as several new sources have been detected displaying the
typical characteristics of SGXBs \citep{wal06}. In most
cases, the sources had not been detected by previous missions due to
high absorption, which renders their spectra very hard.

Moreover, a second major class of SGXBs has emerged. These systems,
instead of being persistent X-ray sources, are detected as transients,
characterised by very short outbursts ($\sim$hours) separated by
long ($\sim$ months) quiescence periods \citep{sgue05,sgue06}. There
are many candidates to belong to this class \citep{sgue06}, but so far
only two objects have been well characterised.

\section{The prototypes}

\subsection{XTE~J1739$-$302 = IGR~J17391$-$3021}

XTE~J1739$-$302 was discovered during an outburst in August 1997
\citep{smi98}. Further observations with {\it RossiXTE}, and
also with {\it ASCA}, showed it to be an unusual kind of transient
\citep{smi06}, with very  
short outbursts. {\it INTEGRAL} observations showed that these
outburst last only a few hours \citep{lut05a,sgue05}. Monitoring of
the Galactic Centre region with {\it INTEGRAL} reveals that flares are
rare, with typical intervals between outbursts
of several months \citep{sgue05}. 

The outbursts start with a very sharp rise (with a timescale $<1$ h)
and sometimes show complex structure, with several flare-like peaks
\citep{lut05a,sgue05}. The X-ray spectrum during the outbursts is
generally very absorbed, though the absorption is variable. Good fits
can be achieved with either a power law with a high-energy cut-off or
a thermal bremsstrahlung model with $kT\sim 20\:$keV
\citep{smi06,lut05a}. 

 The source was not detected during most of an {\it ASCA} pointing in
March 1999 (with an upper
limit $L_{{\rm X}}<10^{33}\:{\rm erg}\,{\rm s}^{-1}$), but went into
outburst at the end of the same observation. 
{\it Chandra} detected the source at a moderate luminosity $L_{{\rm
     X}}\sim10^{34}\:{\rm erg}\,{\rm s}^{-1}$, allowing the identification
     of the counterpart \citep{smi06}. VLT/FORS1 spectra taken in May
     2004 show the 
     counterpart to be an O8\,Iab(f) star, placed at a distance
     $\approx 2.6\:$kpc \citep{neg06}. Interstellar absorption is much
     lower than the absorption implied by X-ray spectral fits. At this
     distance, the luminosity at the peak of the outbursts approaches $L_{{\rm
     X}}\sim10^{36}\:{\rm erg}\,{\rm s}^{-1}$.

\subsection{IGR~J17544$-$2619}
IGR~J17544$-$2619 was discovered by {\it INTEGRAL} on 2003 September 17th,
when it showed two flares lasting $\sim 2\:$h and $\sim 8\:$h
\citep{sun03,greb03}. On 2004 March 8th, it showed a  
complex outburst lasting more than $8\:$h \citep{greb04}. So far, 5
outbursts from this system have been observed by {\it  
  INTEGRAL} \citep{sgue06}.  In
outburst, the spectrum is hard and moderately absorbed, with evidence
for some variation in the amount of absorbing material.

The source was observed by {\it XMM-Newton} on 2003 September 11th and
17th and in both cases seen at $L_{{\rm X}}\sim10^{35}\:{\rm erg}\,{\rm
  s}^{-1}$ \citep{gonz04}, though it was not detected during a
serendipitous observation in March 2003
($L_{{\rm X}}\la2\times10^{32}\:{\rm erg}\,{\rm s}^{-1}$). {\it
  Chandra} observed it on 2004 July 3rd, in a very different state, 
with $L_{{\rm X}}\sim5\times10^{32}\:{\rm erg}\,{\rm
  s}^{-1}$  and displaying a soft spectrum \citep{zand05}.

The counterpart to the source has been unambiguously identified with
the {\it XMM-Newton} and {\it Chandra} positions. The counterpart is
an O9\,Ib supergiant, at a distance of $\sim 3\:$kpc \citep{pel06}.

\section{The fast X-ray transients}

Because of their properties, XTE~J1739$-$302 and IGR~J17544$-$2619
have been termed Supergiant Fast X-ray Transients (SFXTs) by \citep{smi06}.
A number of sources have displayed similar X-ray lightcurves,
characterised by low-level or undetectable quiescence levels and repeated
short X-ray outbursts, among which are IGR J16479$-$4514,
AX~J1749.1$-$2733, SAX J1818.6$-$1703 and AX~J1841.0$-$0536
\citep{sgue05,sgue06}. The last two sources are known to be associated
with massive stars. 

Three other sources have only been observed to flare
once, but are believed to be associated with supergiants. They are
IGR~J08408$-$4503, seen by {\it INTEGRAL} \citep{gotz} and later
detected by {\it Swift} at the position of the O8.5\,Ib supergiant
HD~74194 \citep{kc06}, IGR J16465$-$4507 and AX 1845.0$-$0433. 

Many other systems have shown short outbursts that resemble those of
the known SFXTs \citep[see][]{sgue06,smi06}. Of special interest is
IGR~J11215$-$5952, a source displaying short outbursts every 329~d
\citep{sidoli,smithtel06a}, identified with the B supergiant HD~306414.

\section{New data}

\subsection{IGR J16465$-$4507}

IGR J16465$-$4507 was discovered by {\it INTEGRAL} during an
X-ray flare on 2004 September 7th \citep{lut04}.  A subsequent {\it
  XMM-Newton} observation \citep{lut05b} revealed that the source is a pulsar 
with $P_{{\rm spin}}= 228\:{\rm s}$ and is extremely absorbed ($N_{{\rm
      H}} \sim7\times10^{23}\:{\rm cm}^{-2}$). The {\it XMM-Newton}
      position allows the identification of a counterpart. Our new VLT/FORS1
      spectrum reveals that this object is a B0.5\,Ib
      supergiant. Lacking accurate photometry, we estimate its
      distance at $\sim 8$~kpc.

\subsection{SAX J1818.6$-$1703}

SAX~J1818.6$-$1703 was discovered by {\it BeppoSAX} during a strong
short outburst (with a rise time of $\sim 1\:$h), in March 1998
\citep{zand98}.  {\it INTEGRAL} detected a double-peaked outburst
in September 2003 \citep{gs05} and
two more in October 2003 \citep{sgue05}. Other fast outbursts have
been observed with the ASM on {\it RossiXTE} \citep{sgue05}.

\begin{figure}
\centering
\includegraphics[bb= 50 235 540 565,width=0.95\linewidth]{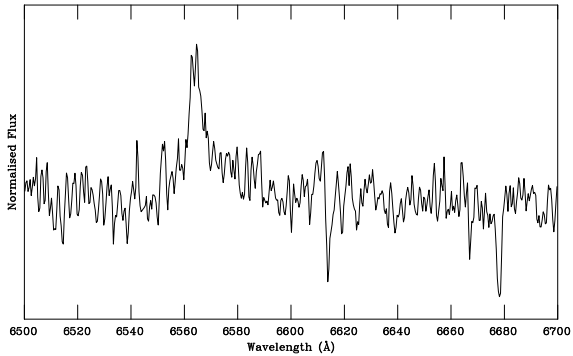}
\caption{Intermediate resolution spectrum of the likely counterpart to
SAX~J1818.6$-$1703, taken with the NOT on June 2006. {\rm H}$\alpha$ is
emission, but the {\rm He}\,{\sc i}~6678\AA\ line is strongly in
absorption. No {\rm He}\,{\sc ii}~6683\AA\ line is visible in its vicinity,
indicating that this object is O9 or later. A lower resolution
spectrum taken with the VLT can be used to constrain the spectral type
to be earlier than B1.
\label{fig:alpha}}
\end{figure}

The X-ray lightcurve of SAX~J1818.6$-$1703 is typical of a SFXT. The
X-ray spectrum is very hard \citep{gs05}. We have found an obscured
supergiant close to the centre of the error circle for
SAX~J1818.6$-$1703 and believe that this is its correct counterpart
\citep{ns06}. A spectrum of this object is shown in
Fig.~\ref{fig:alpha}.

\subsection{AX J1841.0$-$0536}

AX~J1841.0$-$0536 was observed as a violently variable transient by
{\it ASCA} in April 1994 and then again in October 1999
\citep{bam01}. The source showed multi-peaked flares with a sharp rise
(tenfold increase in count-rate over $\sim1\:$h). 
Analysis of the {\it ASCA} data revealed that the source is a pulsar 
with $P_{{\rm spin}}= 4.7\:{\rm s}$. The spectrum can be fit by an
  absorbed power law plus iron line \citep{bam01}. Three flares from
  this source have been observed by {\it INTEGRAL} \citep{sgue06}. Its
  spectrum displays a hard tail up to at least 80~keV.

A {\it Chandra} observation of the field allowed the localisation of
the counterpart to AX~J1841.0$-$0536 \citep{hal04}. Our new VLT/FORS1
spectrum shows 
this object to be a B0.2\,Ib supergiant. 

\subsection{AX 1845.0$-$0433}

AX~1845.0$-$0433 was discovered by {\it ASCA} during a strong flare in
1993. The outburst consisted of a very fast rise (on the order of a
few minutes) followed by a number of peaks during the next few
hours. The spectrum was well fit by an absorbed power law
\citep{yam95}. The {\it ASCA} error circle was studied by
\citet{coe96}, who found 
only one remarkable object, a late O-type supergiant. Our new
VLT/FORS1 spectrum shows this object to be an O9\,Ia supergiant at an
estimated distance of $\sim7\:$kpc.

\section{Persistent transients and outbursting persistent sources}

From the data on {\it INTEGRAL} monitoring of SFXTs
\citep{sgue06}, IGR J16479$-$4514 appears to be the most active fast
transient with 10 short outbursts observed in two years. On the other
hand, IGR J16479$-$4514 is given as a persistent source by
\citet{wal06}. Obviously, the difference between persistent and
transient behaviour may depend on the detection level, but here we
have a source with a detectable quiescence level producing frequent
short outbursts.

Conversely, at least three classical SGXBs have been observed to
undergo flares on a timescale comparable to those seen in SFXTs. These
are Vela X-1, 1E 1145.1-6141 and Cyg X-1 \citep[see][for
references]{smi06}. Similarly, IGR J16418$-$4532, one of the new
persistent sources listed by \citet{wal06}, has also shown an SFXT-like
flare \citep{sgue06}. There are good reasons to believe that
IGR~J16195$-$4945 is a persistent source \citep{tomsick}, but it has
also 
shown a flare \citep{sgue06}. Another source proposed as a SFXT, XTE
J1743-363 \citep{sgue06}, has been observed during the {\it RXTE}
Galactic bulge scan program to show violent variability 
   with many spikes, with a lightcurve which seems intermediate
   between  persistent and SFXT behaviour (C.~Markwardt, private
   communication).

\section{IGR~J11215$-$5952}

IGR~J11215$-$5952 was discovered during a brief
flare in April 2005 \citep{lub05}. Its association with the bright
supergiant  HD~306414 has been confirmed by a {\it SWIFT}
observation \citep{ste06}. Analysis of the  {\it INTEGRAL} lightcurve
for this source
\citep{sidoli} found
 three short ($\sim 2\:$d) outbursts separated by $\sim
 330\:$d. A {\it RossiXTE} ToO observation of the source 330~d
  after the previous outburst resulted in the detection of a new
  2-day-long rather bright ($L_{{\rm X}}\sim10^{36}{\rm erg}\,{\rm s}^{-1}$)
  outburst centred on March 17th, 2006 \citep{smithtel06a} . These
 observations not only confirm the periodicity, but also 
  show that the compact object in IGR~J11215$-$5952 is a NS,
  as an X-ray pulse is almost certainly seen \citep{smithtel06b}.

A new spectrum of the counterpart, HD~306414, was taken with the 1.9-m
telescope at the South African Astronomical Observatory in May 2006
(Fig.~2). We derive a spectral type B0.7\,Ia, in good
agreement with previous works, and estimate the distance at $\sim8$~kpc

\begin{figure*}
\label{fig:blue}
\includegraphics[bb = 170 125 440 625, angle=-90,width=0.95\linewidth]{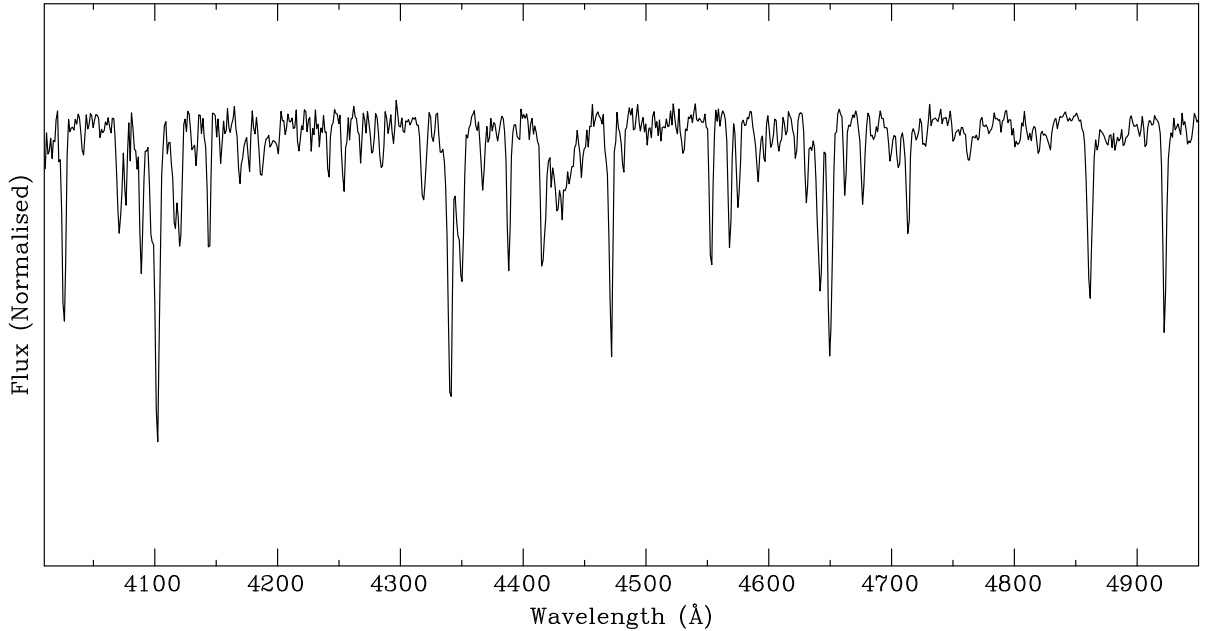}
\caption{Classification spectrum of HD~306414, the optical counterpart
to IGR~J11215$-$5952. The presence of a very weak
{\rm He}\,{\sc ii}~4686\AA\ line puts the spectral type at B0.7\,Ia, rather
than the B1\,Ia derived from lower resolution spectra. The intensity
of {\rm Si}\,{\sc iv}, {\rm Si}\,{\sc iii} and {\rm O}\,{\sc ii}
features shows that 
this is a very luminous supergiant of luminosity class Ia.}
\end{figure*}

IGR~J11215$-$5952 presents strong similarities to SFXTs, including the
short outbursts peaking at $L_{{\rm X}}\sim10^{36}{\rm erg}\,{\rm
s}^{-1}$, separated by long periods of quiescence. Moreover, these
outbursts present irregular lightcurves, containing several
independent flares. On the other hand, these outbursts last on average
2.5~d, as compared to 2\,--\,8~h for more typical SFXTs, and are
separated by a fixed gap. The existence of this periodicity in the
occurrence of the outbursts sets strong constraints on the possible
outburst mechanisms, but, at the same time, sets IGR~J11215$-$5952
apart from other SFXTs.

\section{Conclusions}

A flurry of recent results confirms without any doubt that {\it
INTEGRAL} has found a class of transient
supergiant systems which was not suspected to exist before. However, we
have not found yet any telling characteristic of this new class beyond
the presence of short outbursts. Rather it seems that there is a
continuum in possible behaviours between persistent sources (perhaps
occasionally presenting a flare) and transient systems (at least
sometimes) undetectable in quiescence. The physical mechanism causing
these flares is unknown and the discovery of a recurrent periodic
flaring system, IGR~J11215$-$5952, has only served to cast more doubts
on this issue.

In any case, the number of systems with short outbursts is increasing
steadily and it seems now clear that many more flaring systems may lie
hidden in the Galactic plane.

\section{Note added in proof}

Since this paper was submitted, \citep{sgue07} have reported
repeated flaring activity from AX 1845.0$-$0433 and an accurate
position that identifies it with the O9\,Ia supergiant discussed
here. This object is hence an SFXT.
IGR~J08408$-$4503 has also been confirmed as an SFXT by detection of
further flares \citep{gotz07}. A {\it Chandra} localisation has
identified SAX J1818.6$-$1703 with the obscured supergiant proposed
here \citep{zand06}. IGR~J11215$-$5952 has displayed a new outburst,
allowing a firm 
detection of a pulse period ($P_{{\rm spin}}=186.8\pm0.3$,
\citep{swank}). The outburst, occurring 329 d after the previous one,
was observed to be followed by low flux X-ray emission lasting
$\sim10\:{\rm d}$ \citep{romano07}. 

\section*{Acknowledgements}

IN is a researcher of the programme {\em Ram\'on y Cajal}, funded by
the Spanish Ministerio de Educaci\'on y
Ciencia and the University of Alicante, with partial 
support from the Generalitat Valenciana and the European Regional
Development Fund (ERDF/FEDER).
This research is partially supported by the MEC under
grant AYA2005-00095. During part of this work, IN was a visiting
fellow at the Open University, whose kind hospitality is warmly
acknowledged. This visit was funded by the MEC under grant
PR2006-0310.

Part of the data presented here have been taken using ALFOSC, which is 
owned by the Instituto de Astrof\'{\i}sica de Andaluc\'{\i}a (IAA) and
operated at the NOT under agreement
between IAA and the NBIfAFG of the Astronomical Observatory of
Copenhagen.
This research has made use of the Simbad data base, operated at CDS,
Strasbourg (France).


\end{document}